\documentclass[aps,reprint,twocolumn,superscriptaddress,showpacs,
amsmath,amssymb,footinbib,longbibliography]{revtex4-2}

\usepackage[english]{babel}
\usepackage{amsmath,amssymb}
\usepackage{graphicx,xcolor}
\usepackage[unicode,bookmarksnumbered,breaklinks,colorlinks,linktocpage,
citecolor=blue,linkcolor=darkred,urlcolor=darkblue]{hyperref}
\usepackage{bm}
\usepackage{hyperref}

\definecolor{darkblue}{rgb}{0.0, 0.0, 0.55}
\definecolor{darkred}{rgb}{0.55, 0.0, 0.0}

\begin{document}

%%%%%%%%%%%%%%%%%%%%%%%%%%%%%%%%%%%%%%%%%%%%%%%%%%%%%%%%%%%%%%%%%%%%%%%%%%%%%%%%%%
\title{Thermodynamics of the $S=1/2$ maple-leaf Heisenberg antiferromagnet}

\author{Taras Hutak}
\email{t.hutak@icmp.lviv.ua}
\affiliation{Institute for Condensed Matter Physics,
    National Academy of Sciences of Ukraine,
    Svientsitskii Street 1, 79011 L'viv, Ukraine}

\date{\today}
%%%%%%%%%%%%%%%%%%%%%%%%%%%%%%%%%%%%%%%%%%%%%%%%%%%%%%%%%%%%%%%%%%%%%%%%%%%%%%%%%%
\begin{abstract}
The Heisenberg antiferromagnet on the maple-leaf lattice has recently gathered a great deal of attention. Competition between three nonequivalent bond interactions results in various ground-state quantum phases, the exact dimer-product singlet ground state being among them. The thermodynamic properties of this model are much less understood. We used high-temperature expansion up to the $18$th order to study the thermodynamics of the $S=1/2$ Heisenberg model on the uniform maple-leaf lattice with the ground state exhibiting a six-sublattice $120^{\circ}$ long-range magnetic order. Pad\'{e} approximants allow us to get reliable results up to the temperatures of about $T\approx 0.4$. To study thermodynamics for arbitrary temperatures, we made the interpolation using the entropy method. Based on the analysis of close Pad\'{e} approximants, we find ground-state energy $e_{0}=-0.53064\ldots -0.53023$ in good agreement with numerical results. The specific heat $c(T)$ has a typical maximum at rather low temperatures $T\approx0.379$ and the uniform susceptibility $\chi(T)$ at $T\approx0.49$. We also estimate the value of $\chi(T)$ at zero temperature $\chi_{0}\approx0.05\ldots0.06$. The ground-state order manifests itself in the divergence of the so-called generalized Wilson ratio.
\end{abstract}
%%%%%%%%%%%%%%%%%%%%%%%%%%%%%%%%%%%%%%%%%%%%%%%%%%%%%%%%%%%%%%%%%%%%%%%%%%%%%%%%%%
\maketitle
%%%%%%%%%%%%%%%%%%%%%%%%%%%%%%%%%%%%%%%%%%%%%%%%%%%%%%%%%%%%%%%%%%%%%%%%%%%%%%%%%%
\section{Introduction}
\label{s1}
The Heisenberg antiferromagnetic model on geometrically frustrated lattices is a fruitful playground for studying exotic classical and quantum magnetic phases \citep{knolle2019,broholm2020}. Two-dimensional antiferromagnets are of exceptional interest due to their low dimensionality and strong effects of quantum fluctuations. Usually, quantum fluctuations weaken classical long-range magnetic order (for example, this manifests in the reduction of the sublattice magnetization compared to the classical one). However, for some geometrically frustrated lattices, magnetic ordering can be suppressed completely \citep{richter2004}. The most famous and studied examples of such systems are the triangular-- and kagome--lattice Heisenberg antiferromagnets. The former model exhibits long-range $120^{\circ}$ magnetic order in the ground state \citep{capriotti1999,white2007}, while the latter ground state is now considered to be a quantum spin liquid \citep{yan2011,depenbrock2012}.

Another geometrically frustrated lattice of recent interest is the maple-leaf lattice \citep{betts1995}. Unlike triangular and kagome lattices, it has three non-equivalent bond types: bonds on hexagons, bonds on triangles, and the remaining bonds called dimer bonds, see Fig.~\ref{lattice}. Mainly, the previous studies concerned ground-state behavior, such as phase diagrams and the magnetization process \citep{schulenburg2000,schmalfuss2002,farnell2011,farnell2014,farnell2018,ghosh2022,
ghosh2023,gresista2023,gembe2024,sonnenschein2024,ghosh2024jcm,beck2024,schmoll2024a,ghosh2025}. The most examined line on the parameter space of the general three types of nearest neighbor interactions is the $J\!-\!J_{\text{d}}$ model with equal exchange interactions on triangles and hexagons $J_{\text{h}}\!=\!J_{\text{t}}\!=\!J$. When interaction on the dimer bonds $J_{\text{d}}\!=\!0$, one faces the bounce lattice, and $J_{\text{d}}\!=\!J$ corresponds to the maple-leaf lattice. One of the interesting features of this model is the case $J_{\text{d}}\!\geq\!2J$ when the model hosts the exact dimer-product singlet ground state \citep{farnell2011,ghosh2022} (together with the Shustry--Sutherland lattice \citep{shastry1981}  of coordination number $z\!=\!5$ these are the only two lattices in 2D possibly to host such a state \citep{ghosh2022}).

It was believed earlier that the increasing $J_{\text{d}}$  drives the model to undergo a phase transition from the so-called canted $120^{\circ}$ order at small and intermediate values of $J_{\text{d}}$ to the dimer-singlet product state \citep{farnell2011}. Recent numerical studies by the pseudo-fermion functional renormalization group \citep{gresista2023} and density matrix renormalization group \citep{beck2024} show evidence of an intermediate paramagnetic phase (again quite similar to the phase diagram of the Shasty-Sutherland model \citep{corboz2025}). 

In the present paper, we will focus on the thermodynamic properties of the $S=1/2$ uniform maple-leaf lattice Heisenberg antiferromagnetic model, where all non-equivalent bond interactions are equal. In this peculiar case, the ground state is the classical six-sublattice $120^{\circ}$ magnetic order, or a staggered vector chirality order \citep{haraguchi2018,aguilar-maldonado2025}, significantly weakened by quantum fluctuations and geometrical frustration of lattice. As a result, the model has one of the smallest sublattice magnetizations among the other Archimedean lattices \citep{farnell2014,farnell2018}. Although usually overshone by the fascinating ground-state behavior of the frustrated magnets, finite-temperature properties may also provide valuable insights even into ground-state physics \citep{popp2024,ramirez2025}. We will use the high-temperature expansion (HTE) and the entropy method interpolation to study the finite-temperature properties, such as specific heat $c(T)$ and the uniform susceptibility $\chi(T)$.  

In the broader context, the uniform maple-leaf lattice Heisenberg antiferromagnet ($z=5$, weak magnetic order) can be viewed as an intermediate between triangular ($z=6$, magnetically ordered) and kagome one ($z=4$, no magnetic order). It is interesting to compare the thermodynamic properties of these systems. It has been known for a long time that the kagome-- and triangular--lattice antiferromagnets' specific heat shows an intriguing low-temperature behavior. For the kagome--lattice, both finite-temperature Lanczos \citep{schnack2018}  and entropy method results \citep{bernu2020} confirm that the $c(T)$ has a low-temperature shoulder. On the contrary, the specific heat profile of the triangular lattice is less settled. Tensor network \citep{chen2019} and finite-temperature Lanczos \citep{ulaga2024}  indicate possible two energy scale physics display itself in the two-peak profile of $c(T)$, and entropy method results show one broad maximum \citep{gonzalez2022}.  

Concerning materials, up to date, several natural minerals and synthesized compounds with maple-leaf lattice geometry are known \citep{fennell2011,haraguchi2018,venkatesh2020,haraguchi2021,makuta2021,saha2023,ghosh2024prb,schmoll2024as,aguilar-maldonado2025}. However, studying some of them requires considering the Heisenberg model with higher spin $S>1/2$. The relevant Heisenberg models for $S\!=\!1/2$ materials are complex, and adequate description inevitably involves considering multiple exchange interaction types.      
  
The paper is organized as follows. We begin with a description of the model and methods used in Sec.~\ref{s2}. Results are reported in Sec.~\ref{s3}, followed by the conclusion in Sec.~\ref{s4}.  

%%%%%%%%%%%%%%%%%%%%%%%%%%%%%%%%%%%%%%%%%%%%%%%%%%%%%%%%%%%%%%%%%%%%%%%%%%%%%%%%%%
%%%%%%%%%%%%%%%%%%%%%%%%%%%%%%%%%%%%%%%%%%%%%%%%%%%%%%%%%%%%%%%%%%%%%%%%%%%%%%%%%%
\section{Model and methods}
\label{s2}
\begin{figure}
\includegraphics[width=0.85\columnwidth]{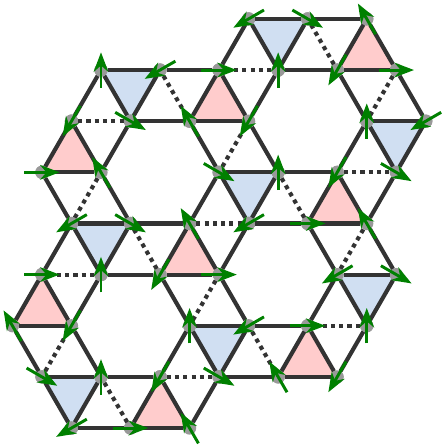}
\caption{Maple-leaf lattice and the classical six-sublattice $120^{\circ}$ ground-state magnetic order. The three nonequivalent interaction bonds are the bonds on hexagons, the bonds on triangles (shaded by blue and red colors),  and the dimer bonds (dotted line). The angle between the classical spin on each bond is $5\pi/6,\pi/3$, and $\pi/2$, respectively \citep{schulenburg2000,schmalfuss2002}.
}
\label{lattice}
\end{figure}

The maple-leaf lattice can be viewed as a 1/7 site-depleted (or a 1/6 bond-depleted) triangular lattice \citep{betts1995}. The lattice has a coordination number $z=5$ with six equivalent sites in a unit cell, and each site belongs to four bonds on triangles and one hexagon bond, see Fig.~\ref{lattice}. We consider the isotropic Heisenberg Hamiltonian on the uniform maple-leaf lattice
\begin{equation}
\mathcal{H}=\sum_{\langle\bm{m}\alpha,\bm{n}\beta\rangle}
\bm{S}_{\bm{m}\alpha}\cdot\bm{S}_{\bm{n}\beta},
\label{eq_ham}
\end{equation}
where the components of spin-1/2 operators $\bm{S}_{\bm{m}\alpha}$ are half of the Pauli matrices. The sum in Eq.~(\ref{eq_ham}) runs over the nearest neighbors' sites of the maple-leaf lattice defined by $\bm{R}_{\bm{m}\alpha}=\bm{R}_{\bm{m}}+\bm{r}_{\alpha}$. Here, 
$\bm{R}_{\bm{m}}=n_{1}\bm{e}_{1}+n_{2}\bm{e}_{2}$ ($\{n_{1},n_{2}\}$ are integers) generates lattice translations, where $\bm{e}_{1}=\frac{1}{2}(5,\sqrt{3}),\bm{e}_{2}=\frac{1}{2}(1,3\sqrt{3})$ are lattice translation vectors, and $\bm{r}_{\alpha}, \alpha=1,\dots,6$ are vectors defining the original position of the six equivalent sites in the unit cell: 
$\bm{r}_{1}=(0,0),\bm{r}_{2}=(1,0),
\bm{r}_{3}=(2,0),\bm{r}_{4}=\frac{1}{2}(1,\sqrt{3}),
\bm{r}_{5}=\frac{1}{2}(3,\sqrt{3}),\bm{r}_{6}=(1,\sqrt{3})$. We set the exchange interaction between the nearest neighbors $J=1$, fixing the energy scale. 

The model (\ref{eq_ham}) has a six-sublattice $120^{\circ}$ long-range magnetic order in the ground state. The angle between classical spins on triangles is $\pi/3$, the angle on hexagons is $5\pi/6$, and finally, on dimer bonds, the angle is $\pi/2$ \citep{schulenburg2000,schmalfuss2002}. Long-range ground-state magnetic ordering in two dimensions leads to the low-temperature behavior of specific heat $c(T)\propto T^{2}$ (we will utilize this to make the interpolation using the entropy method).

%%%%%%%%%%%%%%%%%%%%%%%%%%%%%%%%%%%%%%%%%%%%%%%%%%%%%%%%%%%%%%%%%%%%%%%%%%%%%%%%%%
%%%%%%%%%%%%%%%%%%%%%%%%%%%%%%%%%%%%%%%%%%%%%%%%%%%%%%%%%%%%%%%%%%%%%%%%%%%%%%%%%%

To examine the thermodynamics of the model (\ref{eq_ham}), we use the high-temperature expansion. HTE is a well-established technique for studying finite-temperature properties of quantum Heisenberg magnets \citep{oitmaa2006}. Apart from being restricted only to some temperature range, this method is free from different limitations present in other numerical methods and allows the study of finite-temperature properties of the system in the thermodynamic limit. We used the algorithm of Pierre, Bernu, and Messio \citep{pierre2024} to get the HTE of the logarithm of partition function $\ln Z$ up to 18th order over inverse temperature $\beta=1/T$ \citep{htenote},
\begin{equation}
\ln Z\!=\!\ln 2+\frac{\theta^{2}}{2}+
\frac{1}{6}\bigg[
\sum_{j=0}^{n}\frac{a_{j}\beta^{j}}{(-4)^{j}j!}+
\theta^{2}\sum_{j=0}^{n-1}\frac{b_{j}\beta^{j}}{(-4)^{j}j!}
\bigg],
\label{lnZ}
\end{equation}
where $\theta=\beta h/2$ and $h$ is a longitudinal magnetic field.
The obtained coefficients are reported in Table.~\ref{table_HTE}. Using the raw series (\ref{lnZ}), one can study thermodynamics in zero magnetic field (we will mainly focus on the specific heat $c(T)=\beta^{2}[d^{2}\ln Z/d\beta^{2}]$  and the uniform susceptibility $\chi(T)=1/\beta[d^{2}\ln Z/dh^{2}]$). 

It is well known that the raw series can be improved by using Pad\'{e} approximants $[u,d](\beta)=\mathcal{P}_{\text{u}}(\beta)/\mathcal{Q}_{\text{d}}(\beta)$, where $\mathcal{P}_{\text{u}}(\beta)$ and $\mathcal{Q}_{\text{d}}(\beta)$ are polynomials of $u$ and $d$ order, respectively, and series expansion around $\beta\to0$ of $[u,d](\beta)$ reproduce the original series up to $u+d$ order. Usually, the raw HTE series breaks down at the temperatures of the order of the exchange interaction $T\approx J$. Pad\'{e} approximants allow extending the region of validity to the intermediate temperatures $T\approx J/2$. However, for most models, especially for the frustrated magnets, the low-temperature regime is the most interesting one.  

%%%%%%%%%%%%%%%%%%%%%%%%%%%%%%%%%%%%%%%%%%%%%%%%%%%%%%%%%%%%%%%%%%%%%%%%%%%%%%%%%%
\begin{table*}
\label{table}
\caption{High-temperature expansion series coefficients $a_{j}, b_{j}$ of the 
$\ln Z$ obtained by the algorithm of \citep{pierre2024}.}
%\hskip-1.05cm
\centering
\begin{tabular}{crr}
\hline
\hline
{$n$} & \multicolumn{1}{c}{$a_{j}$} & \multicolumn{1}{c}{$b_{j}$}\\
\hline
$0$&		$0$	&	$0$	  \\
$1$&		$-15$	&	$15$	\\
$2$&		$45$	&	$90$	\\
$3$&		$54$	&	$450$	\\
$4$&	$-1\!~458$	&	$1080$	\\
$5$&	$-9\!~720$	&	$9\!~240$	\\
$6$&	$219\!~240$	&	$587\!~088$	\\
$7$&	$3\!~801\!~168$	&	$8\!~502\!~096$\\	
$8$&	$-67\!~704\!~336$	& $-154\!~746\!~240$\\
$9$&	$-2\!~534\!~315\!~904$	&	$-5\!~666\!~092\!~416$\\
$10$&	$31\!~633\!~948\!~800$	&	$105\!~282\!~266\!~880$\\
$11$&	$2\!~572\!~904\!~353\!~536$	&	$6\!~877\!~641\!~769\!~728$\\
$12$&	$-15\!~715\!~380\!~089\!~088$	&	$-82\!~970\!~518\!~557\!~696$\\
$13$&	$-3\!~721\!~701\!~304\!~488\!~960$	&	
$-10\!~929\!~337\!~956\!~185\!~088$\\
$14$&	$-7\!~224\!~653\!~773\!~089\!~792$	&	
$93\!~115\!~526\!~926\!~682\!~112$\\
$15$&	$7\!~310\!~312\!~980\!~029\!~794\!~304$	&
$24\!~407\!~191\!~222\!~316\!~820\!~480$\\
$16$& 	$92\!~262\!~608\!~432\!~299\!~505\!~664$	&	
$-133\!~859\!~693\!~794\!~581\!~970\!~944$\\
$17$&$-18\!~643\!~625\!~124\!~239\!~891\!~202\!~048$
&$-77\!~080\!~129\!~995\!~065\!~123\!~438\!~592$\\
$18$&$-511\!~312\!~173\!~537\!~287\!~578\!~484\!~736$&\\
\hline
\hline
\end{tabular}
\label{table_HTE}
%\end{ruledtabular}
\end{table*}
%%%%%%%%%%%%%%%%%%%%%%%%%%%%%%%%%%%%%%%%%%%%%%%%%%%%%%%%%%%%%%%%%%%%%%%%%%%%%%%%%%

This obvious limitation of Pad\'{e} approximants can be oversteped using the so-called entropy method to study thermodynamics on the whole temperature range \citep{bernu2001,misguich2005,bernu2015}. This method was applied to the Heisenberg model on several frustrated lattices \citep{misguich2007,schmidt2017,bernu2020,derzhko2020,grison2020,gonzalez2022,hutak2023,hutak2024}. The idea of the entropy method is not to directly interpolate some observable, say, the specific heat, but the entropy $s(e)$ as a function of the internal energy $e$, guided by a knowledge of a low-temperature behavior. In this framework, all observables are defined parametrically on the interval $e\in[e_{0},0]$, where $e_{0}$ is a ground state energy,
\begin{equation}
T=\frac{1}{s^{\prime}(e)},~~~~~
c=-\frac{[s^{\prime}(e)]^{2}}{s^{\prime\prime}(e)}.
\label{T_c_par}
\end{equation}

From the high-temperature expansion of the entropy $s(T)$, one can obtain the expansion of $s(e)=\ln 2 +\sum_{j=2}^{n}s_{j}e^{j}$ around $e_{\infty}=0$  of the same order. To proceed further, the knowledge of the low-temperature behavior of $c(T)$ or the type of excitation is needed. In this paper, we consider only the case of a model with a gapless spectrum. In this scenario, the specific heat has a power-law form at low temperatures $c(T)\propto T^{\alpha}$. Using Eq.~(\ref{T_c_par}) one can immediately find out that such behavior of $c(T)$ leads to entropy $s(e)\propto(e-e_{0})^{\alpha/(1+\alpha)}$.

Finally, to interpolate the entropy between high and low temperatures, we will use the auxiliary function $G(e)$ to remove possible problematic behavior of the $s^{\prime}(e)=\beta$ in the ground state \citep{misguich2005}   
\begin{equation}
\label{eq_sG}	
\begin{aligned}
G(e)&=\frac{\left[s(e)\right]^{\frac{1+\alpha}{\alpha}}}{e-e_0} 
\to 
G_{\rm app}(e){=}\frac{\left(\ln 2\right)^{\frac{\alpha}{1+\alpha}}}{-e_0}
\frac{\mathcal{P}_u(e)}{\mathcal{Q}_d(e)},\\
s_{\rm app}(e)&=\left[\left(e-e_0\right)G_{\rm app}(e)\right]^{\frac{\alpha}{1+\alpha}}.
\end{aligned}
\end{equation}
After the interpolation the entropy $s_{\text{app}}(e)$ can be recovered from the $G_{\text{app}}(e)$. Moreover, the value of the prefactor in $c(T)=AT^{\alpha}$ is $A_{\rm app}=[\alpha^{1+\alpha}/(1+\alpha)^\alpha][G_{\rm app}(e_0)]^\alpha$.
 
In the presence of a small magnetic field, the ground state energy of a gapless model is $e^{h}_{0}=e_{0}-\chi_{0}\frac{h^{2}}{2}$, where $\chi_{0}\equiv\chi(T=0)$ is the value of the uniform susceptibility at zero temperature. Assuming the value of $\chi_{0}$ is known, we can now consider field-dependent entropy $s(e,h)$. After repeating the above-described procedure, we can also study the uniform susceptibility 
$\chi(T)$, 
\begin{equation}
m=\frac{1}{[s^{\prime}(e,h)]}\frac{\partial s(e,h)}{\partial h},~~~~~
\chi=\frac{m}{h}.
\label{m_chi}
\end{equation}

Thus, to study the thermodynamics of a gapless model in the framework of the entropy method, one needs to know the value of the ground-state energy $e_{0}$, the power-law exponent of the specific heat at low temperatures $\alpha$, and the value of the uniform susceptibility at zero temperature $\chi_{0}$. While $e_0$ can be found from the self-consistent procedure based on analysis of close Pad\'{e} approximants and, in our case, $\alpha=2$, the value of $\chi_{0}$ is generally not known.
%%%%%%%%%%%%%%%%%%%%%%%%%%%%%%%%%%%%%%%%%%%%%%%%%%%%%%%%%%%%%%%%%%%%%%%%%%%%%%%%%%

\section{Results}
\label{s3}
\begin{figure}[!]
\includegraphics[width=\columnwidth]{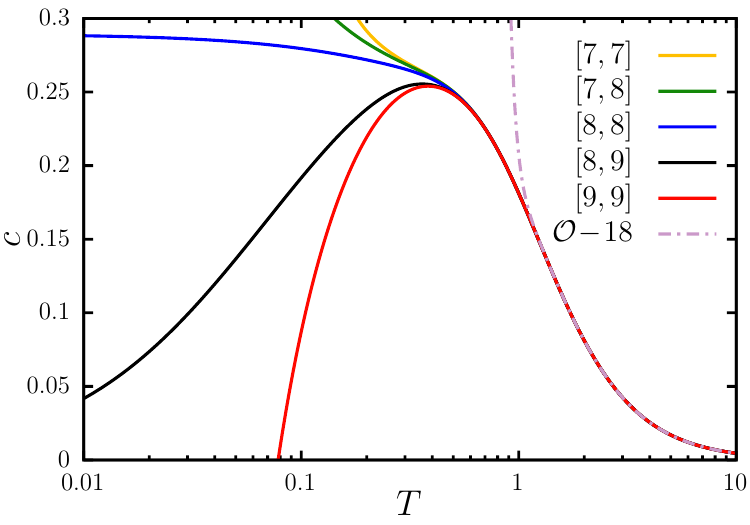}
\includegraphics[width=\columnwidth]{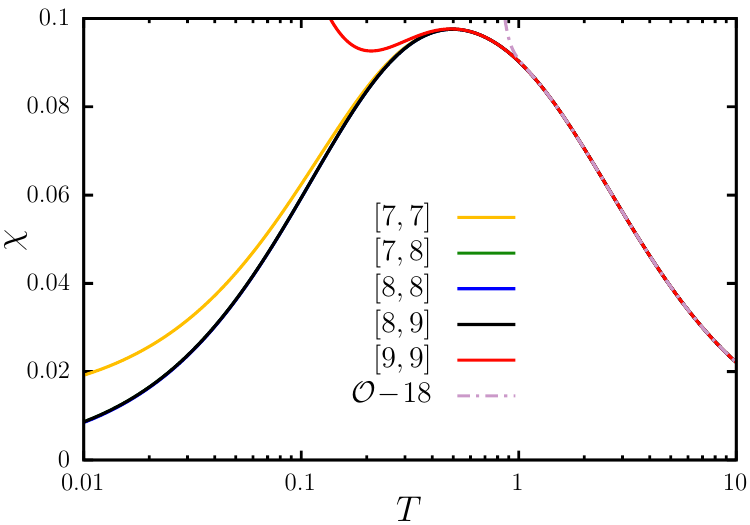}
\caption{Specific heat $c(T)$ (top) and uniform susceptibility $\chi(T)$ (bottom) for the $S=1/2$ antiferromagnetic Heisenberg model on the maple-leaf lattice. Close to diagonal and diagonal Pad\'{e} approximants of HTE order from 14th to 18th are shown. Dashed curves on both panels correspond to the raw series of the 18th order.}
\label{fig_pades}
\end{figure}

We begin a discussion of the thermodynamics of the $S=1/2$ Heisenberg antiferromagnet on the maple-leaf lattice from the analysis 
of the simple Pad\'{e} approximants. Results for the diagonal and close to diagonal Pad\'{e} approximants from the 14th to 18th HTE order are reported in Fig.~\ref{fig_pades}. The raw HTE series of the 18th order for the specific heat $c(T)$ and the uniform susceptibility $\chi(T)$  breaks down at temperatures about $T\approx 1$. By using simple Pad\'{e} approximants, we can reach the temperature $T\approx 0.4$. 

At present HTE order, we can not capture the maximum position of $c(T)$. Pad\'{e} approximants of 18th and 17th order agree to the temperature $T\approx 0.4$, just slightly below the maximum. For the 18th-order Pad\'{e} approximants (the diagonal one shown in Fig.~\ref{fig_pades} and a couple of the off-diagonal Pad\'{e}s), we have $T_{\text{max}}\approx 0.38$ and $c(T_{\text{max}})\approx 0.254$. After integrating the convergent part of the specific heat, one can estimate the upper limit of the ground-state energy value $-\int_{0.4}^{\infty}dT c(T)\approx-0.456$, about 86\% of the $e_{0}$ (see Table~\ref{table_e0}). By utilizing another thermodynamic relation $\Delta s=\int_{0.4}^{\infty}dT c(T)/T\approx0.347\approx \ln2/2$, we can see that down to temperature $T\!=\!0.4$, about half of the total entropy is released. This indicates that the specific heat does not contain an additional low-temperature feature, as one should expect, considering that $c(T)$ recovered almost to its maximum position.

%%%%%%%%%%%%%%%%%%%%%%%%%%%%%%%%%%%%%%%%%%%%%%%%%%%%%%%%%%%%%%%%%%%%%%%%%%%%%%%%%%
%%%%%%%%%%%%%%%%%%%%%%%%%%%%%%%%%%%%%%%%%%%%%%%%%%%%%%%%%%%%%%%%%%%%%%%%%%%%%%%%%%

Contrary to the specific heat, the uniform susceptibility maximum is well captured by the Pad\'{e} approximants $T_{\text{max}}\approx0.497$ and $\chi(T_{\text{max}})\approx0.097$. Approximants from the 15th to 17th order reported in Fig.~\ref{fig_pades} almost coincide.

%%%%%%%%%%%%%%%%%%%%%%%%%%%%%%%%%%%%%%%%%%%%%%%%%%%%%%%%%%%%%%%%%%%%%%%%%%%%%%%%%%
\begin{figure}[!]
\includegraphics[width=\columnwidth]{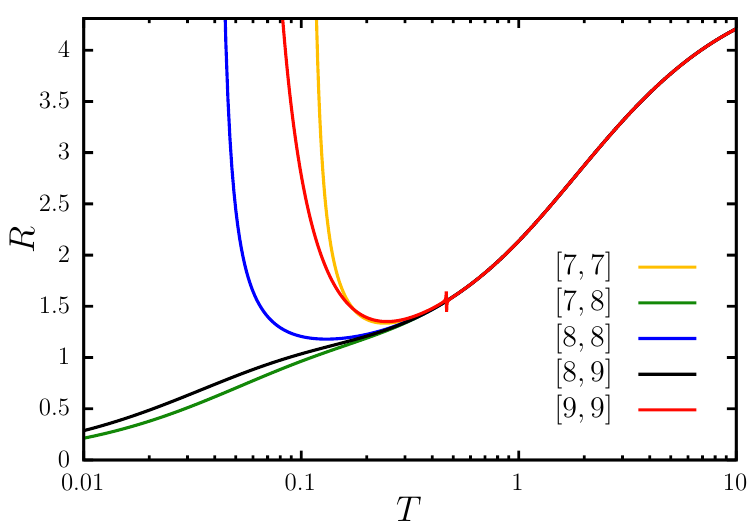}
\caption{Generalized Wilson ratio $R(T)$ Eq.~(\ref{wr}) at different HTE orders.}
\label{fig_wr}
\end{figure}
%%%%%%%%%%%%%%%%%%%%%%%%%%%%%%%%%%%%%%%%%%%%%%%%%%%%%%%%%%%%%%%%%%%%%%%%%%%%%%%%%%

Another useful quantity we can study for the 2D Heisenberg antiferromagnets using HTE is the so-called temperature-dependent generalized Wilson ratio \citep{prelovsek2020prr,prelovsek2020prb}, 
\begin{equation}
R(T)=\frac{4\pi^{2}T\chi(T)}{3s(T)}.
\label{wr}
\end{equation}

This ratio was studied for the finite-size Heisenberg model systems for several 2D lattices \citep{prelovsek2020prr,prelovsek2020prb,seki2020,richter2022,richter2023,ulaga2024}. Contrary to its zero-temperature counterpart in the Fermi-liquid theory \citep{coleman2015}, it contains entropy $s(T)$ instead of the specific heat $c(T)$ in the denominator of Eq.~(\ref{wr}), thus measuring the ratio of the density of the magnetic excitations (with $S^{z}_{\text{total}}>0$) to the density of all excitations (including singlet excitations with $S^{z}_{\text{total}}=0$). 

The high-temperature limit of $R(T)$ for the isotropic Heisenberg model is 
$R\vert_{T\to\infty}=\pi^{2}/(3\ln 2)\approx4.746$. The most interesting is a low-temperature behavior. When low-temperature thermodynamics is dominated by singlet excitations, the generalized Wilson ratio tends to zero $R(T)\propto T^{\eta}\exp(-\Delta/T)$, where $\Delta=\Delta_{\text{t}}-\Delta_{\text{s}}$ is generally the difference between the triplet and singlet gap. Another scenario is when $R(T)$ tends to the finite value $R\vert_{T\to0}\approx 1$, which indicates a gapless spin liquid state with spinon Fermi surfaces \citep{prelovsek2020prr,prelovsek2020prb}. But when the system has a long-range magnetic order in the ground state, as the model at hand, the uniform susceptibility has a finite non-zero value at zero temperature $\chi_{0}>0$, and magnon excitations produce entropy $s(T)\propto T^2$ at low temperatures. This results in the divergence of the generalized Wilson ratio at zero temperature $R(T)\propto T^{-1}$. 

Pad\'{e} approximants of different HTE orders for the generalized Wilson ratio are reported in Fig.~\ref{fig_wr}. Clearly, for the considered model, one can not reach sufficiently low temperatures to get to the characteristic minimum of $R(T)$ (see, for example, numerical studies for Heisenberg antiferromagnets on the square and triangular lattices \citep{richter2022,ulaga2024}) by using a simple Pad\'{e} approximation. However, after a smooth decrease, diagonal Pad\'{e} approximants from the 14th to 18th order reach a minimum at temperatures around $T\!<\!0.3$, starting to increase rapidly after that, thus indicating the presence of the long-range magnetic order. 

Now, let's turn to the entropy method interpolation. As one can see from Eq.~(\ref{eq_sG}), the interpolation requires the knowledge of both the ground-state energy $e_{0}$ and the low-temperature behavior of the specific heat. The low-temperature behavior of the $c(T)$ is governed by the antiferromagnetic spin-wave excitations and is known for the 2D antiferromagnets to be $c(T)\propto T^{2}$ \citep{takahashi1989}. The ground-state energy $e_{0}$ can be found based on the analysis of close (coinciding) Pad\'{e} approximants \citep{bernu2020}. This procedure was applied successfully to several Heisenberg antiferromagnets \citep{ bernu2020,grison2020,gonzalez2022,hutak2024}. Here, we will briefly describe the algorithm used in \citep{hutak2024}. 

From now on, we will treat the ground-state energy $e_{0}$ as a parameter, and varying it with step $10^{-5}$ will check the closeness of the specific heat curves at a given value of $e_{0}$. At the $n$th order, one can build $n+1$ Pad\'{e} approximants. From the start, we will dismiss four Pad\'{e} approximants $[n,0], [n-1,1], [1,n-1], [0,n]$ and form an initial set of $n-3$ curves. The simple Pad\'{e} approximants for the $c(T)$ start disagreeing at $T\approx0.4$, so we evaluate $c(T)$ at a slightly higher temperature $T=0.5$ and remove curves from the initial set which do not reproduce this value. After that, we slowly lower the temperature on each step eliminating the curves that differ from the mean value of the set by 0.001. At the end of this procedure, we end up with a bunch $n_{\text{cP}}$ of very close Pad\'{e} approximants at a given value of $e_{0}$. The greater the number of remaining Pad\'{e} approximants $n_{\text{cP}}$, the closer one gets to the correct ground state value $e_{0}$ \citep{bernu2020}.

%%%%%%%%%%%%%%%%%%%%%%%%%%%%%%%%%%%%%%%%%%%%%%%%%%%%%%%%%%%%%%%%%%%%%%%%%%%%%%%%%%
\begin{figure}[!]
\includegraphics[width=\columnwidth]{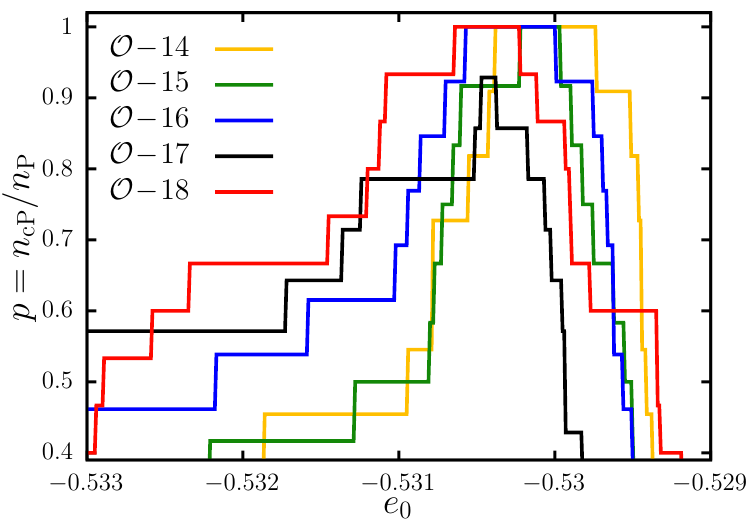}
\caption{Ratio $p = n_{\text{cP}}/n_{\text{P}}$ of the number of close Pad\'{e} approximants $n_{\text{cP}}$ to the number of the initial set considered in the entropy method $n_{\text{P}}$, based on the series of 14th to 18th order as a function of the ground-state value $e_{0}$.}
\label{ncpa}
\end{figure}
%%%%%%%%%%%%%%%%%%%%%%%%%%%%%%%%%%%%%%%%%%%%%%%%%%%%%%%%%%%%%%%%%%%%%%%%%%%%%%%%%%

The results of this search for the HTE order from 14th to 18th are shown in Fig.~\ref{ncpa}. At each order exists a short range of the ground state values where all Pad\'{e} approximants are very close (the 17th order is an exception where all but one curve are very close). We see the tendency of $e_{0}$  very slowly moving to the lower values while increasing the order $n$. At the 18th order, the region where all curves are close is $e_{0}=-0.53064\ldots -0.53023$.

The ground-state energy value was studied by several methods: the exact diagonalization (ED) of the finite $N=36$ system \citep{schmalfuss2002}, the coupled cluster method (CCM) \citep{farnell2011,farnell2014,farnell2018}, the infinite projected entangled-pair state ansatz (iPEPS) \citep{schmoll2024a}, the density matrix renormalization group study (iDMRG) \citep{beck2024}, and the linear spin wave theory (LSWT) \citep{schmalfuss2002}. The entropy method prediction agrees with these findings up to three significant digits, see Table~\ref{table_e0}.  

%%%%%%%%%%%%%%%%%%%%%%%%%%%%%%%%%%%%%%%%%%%%%%%%%%%%%%%%%%%%%%%%%%%%%%%%%%%%%%%%%%
\begin{table}[!]
\caption{Comparison of the ground-state energy values $e_{0}$ obtained by different methods.}
\begin{ruledtabular}
\begin{tabular}{ll}
\hline
ED ($N=36$)  \citep{schmalfuss2002}     &  $-0.5389725$\\
iDMRG \citep{beck2024}		&	$-0.531003325351763$\\
CCM \citep{farnell2018}			&  $-0.53094$                            \\
iPEPS \citep{schmoll2024a}		&  $-0.530359$ \\
entropy method (this work)     &  $-0.53064\ldots -0.53023$                \\
LSWT \citep{schmalfuss2002}    &  $-0.5121574375$                       \\
\end{tabular}
\end{ruledtabular}
\label{table_e0}
\end{table}
%%%%%%%%%%%%%%%%%%%%%%%%%%%%%%%%%%%%%%%%%%%%%%%%%%%%%%%%%%%%%%%%%%%%%%%%%%%%%%%%%%
   
%%%%%%%%%%%%%%%%%%%%%%%%%%%%%%%%%%%%%%%%%%%%%%%%%%%%%%%%%%%%%%%%%%%%%%%%%%%%%%%%%% 
\begin{figure}
\includegraphics[width=\columnwidth]{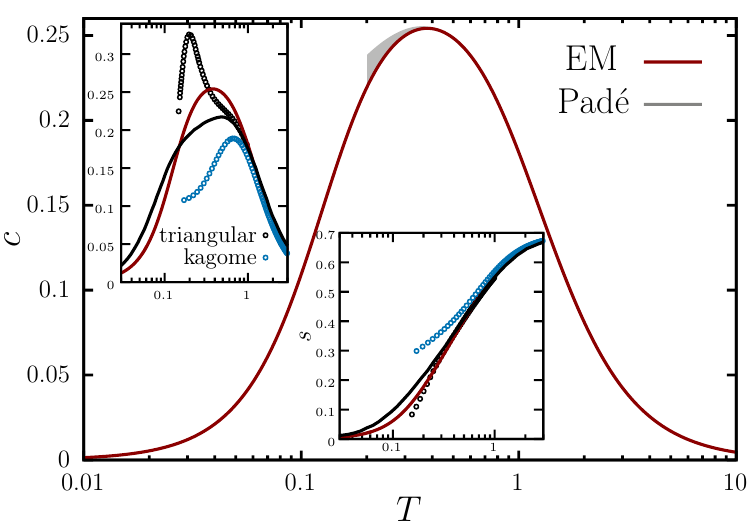}
\includegraphics[width=\columnwidth]{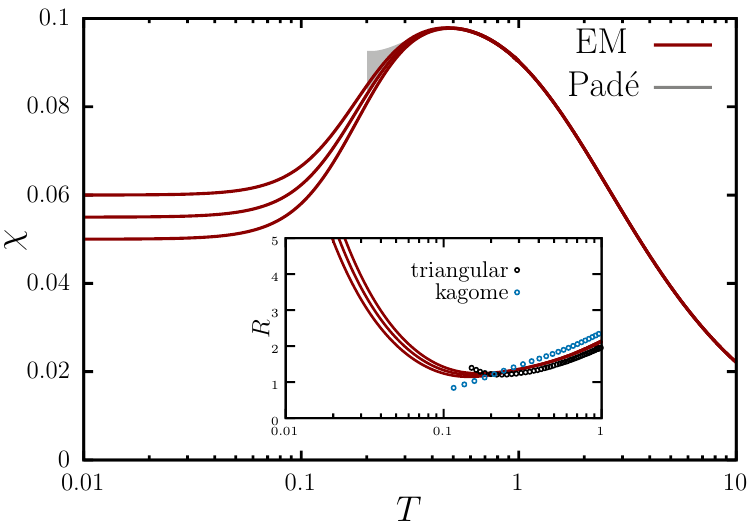}
\caption{Entropy-method results for the specific heat $c(T)$ (top) and the uniform susceptibility $\chi(T)$  (bottom) of the $S\!=\!1/2$ Heisenberg antiferromagnet on the maple-leaf lattice. The insets in the top figure offer a comparison of the maple-leaf $c(T)$ and the entropy $s(T)$ with the finite-system diagonalization for the triangular-- and kagome--lattice models (empty circles, data taken from \citep{ulaga2024} and \citep{schnack2018}, respectively) and entropy method results (solid line, data taken from \citep{gonzalez2022}). The comparison of the generalized Wilson ratio $R(T)$ to the triangular- and kagome-lattice is shown in the bottom inset, assuming plausible values $\chi_{0}\!=\!0.05, 0.055, 0.06$. On both panels, the shaded gray area corresponds to the difference between simple Pad\'{e} approximants $[8,9]$ and $[9,9]$.}
\label{c_chi_EM}
\end{figure}
%%%%%%%%%%%%%%%%%%%%%%%%%%%%%%%%%%%%%%%%%%%%%%%%%%%%%%%%%%%%%%%%%%%%%%%%%%%%%%%%%%
   
We now move on to the thermodynamic properties. We utilize ground-state values $e_{0}$ and use Eqs. (\ref{T_c_par}) and (\ref{m_chi}) to study the specific heat $c(T)$, uniform susceptibility $\chi(T)$, and generalized Wilson ratio $R(T)$. These results are reported in Fig.~\ref{c_chi_EM} (we use $e_{0}=-0.53044$, and all curves correspond to the $[9,9]$ Pad\'{e} approximants). 

The specific heat has one maximum at $T_{\text{max}}\approx0.379$ and $c(T_{\text{max}})\approx0.254$, located at slightly lower temperatures than accessible by only using simple Pad\'{e} approximants. In the framework of the entropy method, one can also calculate the prefactor in the low-temperature power-law of $c(T)$, in our case, $c(T)\approx13.4T^2$. The dependence of the specific heat on the ground-state energy value and the HTE order is very weak for the range of $e_{0}$ where $p=p_{\text{max}}$ (the obtained curves are almost indistinguishable).

The study of uniform susceptibility requires knowledge of its value at zero temperature $\chi_{0}$. Unfortunately, there is no available data on $\chi_{0}$ in the literature. We can again use it as a parameter to correctly reproduce the maximum position of $\chi(T)$, captured by simple Pad\'{e} approximants. Guided by this knowledge, we found that $\chi_{0}$ belongs in the range $\chi_{0}\approx0.05...0.06$. The other choices of its value can spoil the correct intermediate-temperature behavior of the uniform susceptibility.    

The generalized Wilson ratio shows typical behavior for the ordered antiferromagnetic systems \citep{richter2022,ulaga2024}. Slowly decreasing from its high-temperature value, it reaches a minimum at the temperature around $T\approx0.16$ (depending on the chosen $\chi_{0}$ value), starting after that increasing rapidly. $R(T)$ depends very slightly on the particular choice of $\chi_{0}$, see the inset in the bottom panel of Fig.~\ref{c_chi_EM} data for $\chi_{0}=0.05, 0.055, 0.06$. 

Finally, it is worth putting our studies in the context of two other frustrated systems mentioned in the Introduction, triangular-- and kagome--lattice Heisenberg antiferromagnets, see the insets in Fig.~\ref{c_chi_EM}. The specific heat behavior of the maple-leaf antiferromagnet is quite similar to the one obtained by the entropy method for the triangular lattice \citep{gonzalez2022} (likely, the additional pronounced low-temperature maximum for the $N=36$ system is the finite-size effect). Although instead of a broad maximum located at $T\approx0.48$ for the triangular lattice \citep{gonzalez2022}, potentially signaling the existence of two energy scales in the excitation spectra, the maple-leaf $c(T)$ shows one typical maximum at lower temperatures. This indicates that the characteristic energy scale for the maple-leaf lattice is smaller than the triangular lattice one. 

It was noticed in \citep{elstner1993,elstner1994} that the triangular lattice Heisenberg antiferromagnet retains entropy down to low temperatures (contrary to a square lattice antiferromagnet). In this regard, the behavior of the maple-leaf and the triangular lattices is again quite similar. Both models retain entropy at rather low temperatures (see the inset in the top panel of Fig.~\ref{c_chi_EM}). However, the maple-leaf model releases entropy faster. Lastly, we also comment on the uniform susceptibility and the Wilson ratio of these models. There is no reliable data for $\chi(T)$ at the very low temperatures: the intermediate-temperature studies for the triangular lattice \citep{kulagin2013,bruognolo2017} show that $\chi(T)$ is also similar to the maple-leaf one, the zero-temperature value of the susceptibility obtained in the large--$S$ expansion $\chi_{0}\!=\!0.07$ \citep{chubukov1994}, and the high-temperature study estimation $\chi_{0}\!\approx\!0.05$ \citep{elstner1993,elstner1994} close to our predictions for the maple-leaf lattice. This leads to a similar behaviour of the Wilson ratio at intermediate temperatures. The minimum of $R(T)$ for the maple-leaf lattice is at $T\!\approx\!0.16$, and the triangular lattice at a bit higher temperature $T\!\approx\!0.2$ \citep{ulaga2024}. We would like to note here the noticeable difference in the thermodynamic behavior between the frustrated, although ordered models (maple-leaf and triangular), and the kagome lattice antiferromagnet. For the latter case, the $c(T)$ shows two distinct energy scales, $s(T)$ is released significantly slower, and $R(T)$ tends to vanish at lower temperatures \citep{ulaga2024}. 
\section{Conclusion}
\label{s4}
In the present paper, we studied the thermodynamics of the $S=1/2$ Heisenberg antiferromagnet on the uniform maple-leaf lattice based on the high-temperature series expansion up to the 18th order obtained using the algorithm of \citep{pierre2024}. We discuss the thermodynamic properties of the specific heat $c(T)$, the uniform susceptibility $\chi(T)$, and the generalized Wilson ratio $R(T)$. Simple Pad\'{e} approximation allows us to reach the temperatures $T\approx0.4$. At this temperature, we can confidently determine the maximum position of the uniform susceptibility $T\approx0.49$. However, the maximum position of the specific heat at this temperature is not reached yet. 

To study finite-temperature properties at arbitrary temperatures, we used the entropy method. Based on the exact knowledge of the low-temperature behavior of the specific heat, we made an interpolation between high and low temperatures. Our main findings are the following. Using the procedure of finding ground-state energy based on the analysis of close (almost coinciding) Pad\'{e} approximants, we found $e_{0}$ to be in a very narrow region $e_{0}=-0.53064\ldots -0.53023$. The obtained $e_{0}$ value is in good agreement with recent numerical studies. The specific heat shows one maximum located at $T\approx0.379$ just slightly lower than the temperatures reached by simple Pad\'{e} approximants. We also predict the value of the uniform susceptibility at zero temperature $\chi_{0}\approx0.05\ldots0.06$. The thermodynamic behavior of the maple-leaf antiferromagnet is similar to the triangular lattice one, however, with a lower characteristic temperature energy scale.
     
\section*{Acknowledgements}
I acknowledge the discussions with Johannes Richter at the early stages of the work and am grateful to Oleg Derzhko and Dmytro Yaremchuk for the fruitful discussions and for reading the manuscript. I thank the authors of the paper \citep{pierre2024} for making their algorithm open, and I am particularly grateful to Laura Messio for the valuable discussion about running the algorithm. I also thank Pratyay Ghosh for sharing the iDMRG ground-state energy value from \citep{beck2024}.

The work was supported by the fellowship of the National Academy of Sciences of Ukraine for young scholars, and by the Projects of research works of young scientists of the National Academy of Sciences of Ukraine (project \# 29-04/18-2023, Frustrated quantum magnets at finite temperatures), as well as by the EURIZON project (\#3025 "Frustrated quantum spin models to explain the properties of magnets over wide temperature range''), which is funded by the European Union under grant agreement No.~871072.             

I am grateful to the Armed Forces of Ukraine for their protection during this study.

\bibliography{maple-leaf}

\end{document}